\documentclass[aps,prd,draft,showpacs]{revtex4}

\begin{document}

\title{Properties of the vacuum expectation values in $R_{\xi }$ and general gauge}
\author{Chungku Kim}
\affiliation{Department of Physics, College of Natural Science, Keimyung
University, Daegu 705-701, KOREA }
\date{\today}

\begin{abstract}
We have investigated the renormalization group evolution and the
perturbative expansion of the vacuum expectation value(VEV) of the Abelian
Higgs model both in the $R_{\xi }$ and the general gauge which extrapolate
between the Fermi and the $\overline{R_{\xi }\text{ }}$ gauge. In $R_{\xi }$
gauge, the gamma function of the VEV was different from that of the scalar
field. On the contrary, the gamma function of the VEV was same as that of
the scalar field in general gauge. Both in $R_{\xi }\text{ }$ and $\overline{%
R_{\xi }\text{ }}$ gauge, the two-loop VEVs have an IR divergences in Landau
gauge($\xi =0$) and these IR divergence do not occur when $\xi \neq 0$.
\end{abstract}
\maketitle



\smallskip Recently, the renormalization properties of the VEV in
spontaneous broken gauge symmetry with $R_{\xi }$ gauge\cite{R-ksi} were
investigated\cite{VEV-1}\cite{VEV-2}. It was shown that the gamma function
of the VEV ($\gamma _{v}$) have additional contributions originating from
the scalar background field and as a result, $\gamma _{v}$ was found to be
different from that of the scalar field ($\gamma _{\phi }$). Moreover, it
was found that perturbative expansion of the effective action has an IR
divergence in the Landau($\xi =0$) gauge\cite{IR-1}\cite{IR-2}. The
advantage of the R$_{\xi }$ gauge is that the mixing term between the gauge
and the Goldstone fields in the quadratic part of the Lagrangian of the
spontaneously broken symmetry phase cancels out with the gauge fixing term
and hence the propagators between different fields do no exist. The $%
\overline{R_{\xi }\text{ }}$ gauge introduced by Kastening for some other
reason\cite{Kas} also has this property. In this paper, by using the
Abelian-Higgs model, we will investigate the renormalization group(RG)
function of the VEV\ from the RG invariance of the effective potential both
in $R_{\xi }$ and the general gauge which extrapolates between the Fermi and
the $\overline{R_{\xi }\text{ }}$ gauge and then we will calculate the
perturbative expansion of the VEV from the no-tadpole condition in order to
test the IR divergence in this gauge.

We start with the Euclidean Lagrangian density of the Abelian Higgs model given by 
\begin{equation}
L_{AH}(\Phi _{1},\Phi _{2},A_{\mu })=\frac{1}{4}F_{\mu \nu }F_{\mu \nu }+%
\frac{1}{2}(\partial _{\mu }\Phi _{1}+gA_{\mu }\Phi _{2})^{2}+\frac{1}{2}%
(\partial _{\mu }\Phi _{2}-gA_{\mu }\Phi _{1})^{2}+\frac{1}{2}m^{2}(\Phi
_{1}^{2}+\Phi _{2}^{2})+\frac{\lambda }{24}(\Phi _{1}^{2}+\Phi
_{2}^{2})^{2}+counter\text{ }terms,
\end{equation}
with 
\begin{equation}
F_{\mu \nu }=\partial _{\mu }A_{\nu }-\partial _{\upsilon }A_{\mu }.
\end{equation}
where $f(A_{\mu },\Phi )$ is the gauge-fixing function. When $m^{2}>0,$ $%
L_{AH}(\Phi _{1},\Phi _{2},A_{\mu })$ have a local $O(2)$ symmetry given by 
\begin{equation}
\delta _{\theta }A_{\mu }=\partial _{\mu }\theta ,\text{ }\delta _{\theta
}\Phi _{1}=\theta \Phi _{2},\text{ }\delta _{\theta }\Phi _{2}=-\theta \Phi
_{1},
\end{equation}
and when $m^{2}<0,\Phi _{1}$ develops a VEV and the $O(2)$ symmetry is
spontaneously broken. The Lagrangian density in the broken phase can be
obtained by substituting $\Phi _{1}$ with $H+v$ where the Higgs field $H$
have a vanishing VEV.

\subsection{R$_{\xi }$ gauge}

The gauge-fixing function of the R$_{\xi }$ gauge is given in the broken
symmetry phase of the theory as 
\begin{equation}
f(A_{\mu },\Phi _{2})=\partial _{\mu }A_{\mu }-\xi gv\Phi _{2},
\end{equation}
and the total Lagrangian in the broken symmetry phase is given by 
\begin{equation}
L_{BS}(H,\Phi _{2},A_{\mu })=L_{AH}(H+v,\Phi _{2},A_{\mu })+\frac{1}{2\xi }%
f(A_{\mu },\Phi _{2})^{2}+\overline{c}\frac{\delta f(A_{\mu }^{\theta },\Phi
_{2}^{\theta })}{\delta \theta }c+counter-terms,
\end{equation}
where $H$ is the Higgs field which have vanishing VEV and $c$ and $\overline{%
c}$ are the ghosts. As noted above, the quadratic term $gv\Phi _{2}A_{\mu }$
in the $L_{AH}(H+v,\Phi _{2},A_{\mu })$ cancels out with that in $\frac{1}{%
2\xi }f(A_{\mu },\Phi _{2})^{2}$ and the propagator between $\Phi _{2}$ and $%
A_{\mu }$ field does not exists in the total Lagrangian in the broken
symmetry phase $L_{BS}(H,\Phi _{2},A_{\mu }).$ In order to obtain the
one-loop effective potential, we shift $H$ with $H+\phi $ where $\phi $ is
the classical field to obtain $\overline{L}$ defined as\cite{Jackiw} 
\begin{equation}
\overline{L}=L_{BS}(H+\phi ,\Phi _{2},A_{\mu })-L_{BS}(\phi ,\Phi
_{2},A_{\mu })-\left[ \frac{\delta L_{BS}(H,\Phi _{2},A_{\mu })}{\delta H}%
\right] _{H=\phi }H.
\end{equation}
Then the quadratic parts of $\overline{L}$ in momentum space are given as 
\begin{equation}
\frac{1}{2}HD_{H}^{-1}H+\frac{1}{2}( 
\begin{array}{ll}
\Phi _{2} & A_{\mu }
\end{array}
)\left( 
\begin{array}{ll}
D_{G}^{-1} & B_{\nu } \\ 
B_{\mu } & D_{\mu \nu }^{-1}
\end{array}
\right) \left( 
\begin{array}{l}
\Phi _{2} \\ 
A_{\nu }
\end{array}
\right) -\overline{c}D_{g}^{-1}c,
\end{equation}
where 
\begin{equation}
D_{H}^{-1}=p^{2}+m_{H}^{2},
\end{equation}
\begin{equation}
D_{G}^{-1}=p^{2}+m_{G}^{2}+\xi g^{2}v^{2},
\end{equation}
\begin{equation}
D_{\mu \nu }^{-1}=(p^{2}+m_{A}^{2})(\delta _{\mu \nu }-\frac{p_{\mu }p_{\nu }%
}{p^{2}})+(\frac{p^{2}}{\xi }+m_{A}^{2})\frac{p_{\mu }p_{\nu }}{p^{2}},
\end{equation}
\begin{equation}
B_{\mu }=g\phi \text{ }p_{\mu },
\end{equation}
and 
\begin{equation}
D_{g}^{-1}=p^{2}+m_{g}^{2},
\end{equation}
with 
\begin{equation}
m_{H}^{2}=m^{2}+\frac{\lambda }{2}(\phi +v)^{2},m_{G}^{2}=m^{2}+\frac{%
\lambda }{6}(\phi +v)^{2},m_{A}=g(\phi +v),m_{g}^{2}=\xi gvm_{A}.
\end{equation}
Let us define $X_{\mu \nu }^{-1}$ as 
\begin{equation}
X_{\mu \nu }^{-1}\equiv D_{\mu \nu }^{-1}+B_{\mu }D_{G}B_{\nu }.
\end{equation}
By using Eqs.(9),(10) and (11) , we obtain 
\begin{equation}
X_{\mu \nu }^{-1}=(p^{2}+m_{A}^{2})(\delta _{\mu \nu }-\frac{p_{\mu }p_{\nu }%
}{p^{2}})+\frac{D(p^{2})}{m_{G}^{2}+\xi m_{A}^{2}}\frac{p_{\mu }p_{\nu }}{%
\xi p^{2}},
\end{equation}
where 
\begin{equation}
D(p^{2})=p^{4}+(m_{G}^{2}\text{ }+2\xi gvm_{A})p^{2}+\xi
m_{A}^{2}(m_{G}^{2}+\xi g^{2}v^{2})\equiv (p^{2}+m_{+}^{2})(p^{2}+m_{-}^{2}).
\end{equation}
Then the one-loop effective potential $V_{1}$ is given by \cite{Tye} 
\begin{equation}
V_{1}=-\frac{\hbar }{2}Tr\ln D_{H}^{-1} -\frac{\hbar }{2}Tr\ln D_{G}^{-1}
-\frac{\hbar }{2}Tr\ln X^{-1}+\hbar \text{ }Tr\ln D_{g}^{-1}.
\end{equation}
By performing the one-loop momentum integral in $D\equiv 4-2\varepsilon $
dimension\cite{integral}, we can obtain the renormalized one-loop effective
action in the MS scheme as 
\begin{eqnarray}
V &=&V_{0}+V_{1}=\frac{1}{2}m^{2}(\phi +v)^{2}+\frac{1}{24}\lambda (\phi
+v)^{4}+\frac{\hbar }{16\pi ^{2}}\{\frac{1}{4}m_{H}^{4}(\overline{\ln }\text{
}m_{H}^{2}-\frac{3}{2})+\frac{1}{4}m_{+}^{4}(\overline{\ln }\text{ }%
m_{+}^{2}-\frac{3}{2})+\frac{1}{4}m_{-}^{4}(\overline{\ln }\text{ }m_{-}^{2}-%
\frac{3}{2})  \nonumber \\
&&+\frac{3}{4}m_{A}^{2}(\overline{\ln }m_{A}^{2}-\frac{5}{6})-\frac{1}{2}%
m_{g}^{4}(\overline{\ln }\text{ }m_{g}^{2}-\frac{3}{2})\}.
\end{eqnarray}
where $\overline{\ln }X\equiv \ln \frac{X}{4\pi \mu ^{2}}+\gamma .$ The
effective potential is independent of the renormalization mass scale $\mu $
and should satisfy the RG equation\cite{RG} 
\begin{equation}
\mu \frac{dV}{d\mu }=(\mu \frac{\partial }{\partial \mu }+\beta _{\lambda }%
\frac{\partial }{\partial \lambda }+\beta _{g}\frac{\partial }{\partial g}%
+\beta _{m^{2}}\frac{\partial }{\partial m^{2}}+\gamma _{\phi }\phi \frac{%
\partial }{\partial \phi }+\gamma _{v}v\frac{\partial }{\partial v})V=0.
\end{equation}
By using the RG functions in the MS scheme\cite{RG-1}\cite{RG-2} 
\begin{eqnarray}
\beta _{\lambda } &=&\mu \frac{d\lambda }{d\mu }=\frac{\hbar }{(4\pi )^{2}}(%
\frac{10}{3}\lambda ^{2}-12\lambda g^{2}+36g^{4})+\cdot \cdot \cdot , 
\nonumber \\
\beta _{g} &=&\mu \frac{dg}{d\mu }=\frac{\hbar }{(4\pi )^{2}}\text{{}}\frac{%
g^{3}}{3}+\cdot \cdot \cdot ,  \nonumber \\
\beta _{m^{2}} &=&\mu \frac{dm^{2}}{d\mu }=\frac{\hbar }{(4\pi )^{2}}(\frac{4%
}{3}\lambda -6g^{2})m^{2}+\cdot \cdot \cdot ,  \nonumber \\
\gamma _{\phi } &=&\frac{\mu }{\phi }\frac{d\phi }{d\mu }=\frac{\hbar }{%
(4\pi )^{2}}(3-\xi )g^{2}+\cdot \cdot \cdot .
\end{eqnarray}
Then, in order to satisfy the RG equation given in Eq.(19), we should have 
\begin{equation}
\gamma _{v}=\frac{\mu }{v}\frac{dv}{d\mu }=\frac{\hbar }{(4\pi )^{2}}(3+\xi
)g^{2}+\cdot \cdot \cdot \cdot ,
\end{equation}
which agrees with \cite{VEV-1}\cite{VEV-2}. The perturbative expansion of
the VEV $v=v_{0}+\hbar v_{1}+\hbar ^{2}v_{2}+\cdot \cdot \cdot $ can be
obtained from the no-tadpole condition 
\begin{equation}
\left[ \frac{\delta V}{\delta \phi }\right] _{\phi =0}=0,
\end{equation}
as 
\begin{equation}
\left[ \frac{\delta V_{0}}{\delta \phi }\right] _{\phi =0,\text{ }v=v_{0}}=0,
\end{equation}
and 
\begin{equation}
v_{1}\left[ \frac{\delta ^{2}V_{0}}{\delta \phi \delta v}\right] _{\phi =0,%
\text{ }v=v_{0}}+\left[ \frac{\delta V_{1}}{\delta \phi }\right] _{\phi =0,%
\text{ }v=v_{0}}=0.
\end{equation}
By substituting the effective potential given in Eq.(18) into Eqs.(22) and
(23), we obtain the perturbative expansion of the VEV up to one-loop as $%
v_{0}=\sqrt{-\frac{6m^{2}}{\lambda }}$ and 
\begin{eqnarray}
v_{1} &=&-\frac{3}{\lambda v_{0}^{2}}\left[ \frac{\delta V_{1}}{\delta \phi }%
\right] _{\phi =0,\text{ }v=v_{0}}=\frac{1}{32\pi ^{2}m^{2}}[\text{ }\frac{1%
}{2}m_{H}^{2}\frac{\partial m_{H}^{2}}{\partial \phi }(\overline{\ln }\text{
(}m_{H}^{2})-1)+\frac{1}{2}m_{+}^{2}\frac{\partial m_{+}^{2}}{\partial \phi }%
(\overline{\ln }\text{ (}m_{+}^{2})-1)  \nonumber \\
&&+\frac{1}{2}m_{-}^{2}\frac{\partial m_{-}^{2}}{\partial \phi }(\overline{%
\ln }\text{ (}m_{-}^{2})-1)+\frac{3}{2}m_{A}^{2}\frac{\partial m_{A}^{2}}{%
\partial \phi }(\overline{\ln }m_{A}^{2}-\frac{1}{3})-m_{g}^{2}\frac{%
\partial m_{g}^{2}}{\partial \phi }(\overline{\ln }\text{ }m_{g}^{2}-1)\text{
]}_{\phi =0,\text{ }v=v_{0}}.
\end{eqnarray}
From Eq.(16) one can see that by writing $m_{\pm }^{2}=p\pm \sqrt{q}$, both $%
q\ $and $\frac{\partial q}{\partial \phi }$ goes to zero in the limit $\phi
\rightarrow 0$ and $v\rightarrow v_{0}$ and hence $\frac{\partial m_{\pm
}^{2}}{\partial \phi }$ as well as $v_{1}$ have finite limit in $R_{\xi }$
gauge. Now, let us consider IR divergence of the VEV in the case of the
two-loop effective potential\cite{two-1}\cite{two-2} where the term which
can give the IR divergence in effective potential $V_{2}$ and the VEV in
case of the Landau gauge $(\xi = 0 )$ is\cite{IR-1}\cite{IR-2} 
\begin{equation}
m_{X}^{2}\text{ }(m_{G}^{2}+{\xi }g^{2}v^{2})\text{ }\overline{\ln }m_{X}%
\text{ }\overline{\ln }(m_{G}^{2}+{\xi }g^{2}v^{2}),
\end{equation}
coming from the two-loop Feynman diagram 
\begin{equation}
\begin{picture}(80,20) \put(24,10){\circle{24}} \put(48,10){\circle{24}}
\put(64,10){X} \put(0,10) {G} \end{picture},
\end{equation}
where G is the Goldstone boson and X can be Higgs, gauge boson or ghost.
Then, by noting that the mass of the Goldstone boson given as $m_{G}^{2}+{%
\xi }g^{2}v^{2}$ in the $R_{\xi }$ gauge does not vanish in the limit $%
v\rightarrow v_{0}$ and $\phi \rightarrow 0$ when $\xi \neq 0$, the
corresponding two-loop VEV obtained $[$ $\frac{\partial V_{2}}{\partial \phi 
}$]$_{\phi =0,\text{ }v=v_{0}}$does not have an IR divergence in $R_{\xi }$
gauge as long as $\xi \neq 0.$

\subsection{The general gauge}

The general gauge is defined as 
\begin{equation}
f(\Phi _{1},\Phi _{2},A_{\mu })=\partial _{\mu }A_{\mu }-u\xi g\Phi _{1}\Phi
_{2},
\end{equation}
which becomes the Fermi gauge when $u=0$ and $\overline{R_{\xi }}$ gauge
when $u=1.$ The resulting Lagrangian in the symmetric phase is given by\cite
{Kas} 
\begin{equation}
L_{SYM}(\Phi _{1},\Phi _{2},A_{\mu })=L_{AH}(\Phi _{1},\Phi _{2},A_{\mu })+%
\frac{1}{2\xi }f(\Phi _{1},\Phi _{2},A_{\mu })^{2}+\overline{c}(-\partial
^{2}+u\xi g(\Phi _{1}^{2}-\Phi _{2}^{2}))c+counter-terms.
\end{equation}
When $m^{2}<0,$ the $O(2)$ symmetry breaks down spontaneously and we
substitute $\Phi _{1}$ with $H+v$ where $H$ have vanishing VEV to obtain the
Lagrangian density in the broken phase $L_{BS}(H,\Phi _{2},A_{\mu },v)$ as 
\begin{equation}
L_{BS}(H,\Phi _{2},A_{\mu },v)=\left[ L_{SYM}(\Phi _{1},\Phi _{2},A_{\mu
})\right] _{\Phi _{1}\rightarrow H+v}.
\end{equation}
Recently, we have shown that if the Lagrangian in the symmetric phase $%
L_{SYM}(\Phi _{1},\Phi _{2},A_{\mu })$ and that in broken symmetry phase $%
L_{BS}(H,\Phi _{2},A_{\mu },v)$ is related as in Eq.(30), we can prove that $%
\gamma _{v}=\gamma _{\phi }$ and the RG functions of the broken symmetry
phase is same as that of the RG functions of the symmetric phase\cite{Kim1} 
\cite{Kim2}. Note that in case of the $R_{\xi }$ gauge, the gauge fixing and
the ghost terms given in Eq.(5) does not have this relation. In case of the $%
\overline{R_{\xi }}$ gauge $(u=1)$, since the $g\Phi _{1}(\partial _{\mu
}\Phi _{2})A_{\mu }$ term in the $L_{AH}(\Phi _{1},\Phi _{2},A_{\mu })$
cancels out with that in $\frac{1}{2\xi }f(\Phi _{1},\Phi _{2},A_{\mu })^{2}$
and only the $g\Phi _{2}(\partial _{\mu }\Phi _{1})A_{\mu }$ term remains in 
$L_{SYM}(\Phi _{1},\Phi _{2},A_{\mu }),$ the quadratic part of $%
L_{BS}(H,\Phi _{2},A_{\mu },v)$ do not have mixing term between $\Phi _{2}$
and $A_{\mu }$ field as in case of the ${R_{\xi }}$ gauge. In order to
obtain the potential in the broken symmetry phase, we shift $H\rightarrow $ $%
H+$ $\phi $ where $\phi $ is a classical field to obtain $\overline{L}$
defined as 
\begin{equation}
\overline{L}=L_{BS}(H+\phi ,\Phi _{2},A_{\mu },v)-L_{BS}(\phi ,\Phi
_{2},A_{\mu },v)-\left[ \frac{\delta L_{BS}(H,\Phi _{2},A_{\mu },v)}{\delta H%
}\right] _{H=\phi }H.
\end{equation}
Then the quadratic parts of $\overline{L}$ in momentum space is given by 
\begin{equation}
\frac{1}{2}HD_{H}^{-1}H+\frac{1}{2}(
\begin{array}{ll}
\Phi _{2} & A_{\mu }
\end{array}
)\left( 
\begin{array}{ll}
D_{uG}^{-1} & B_{u\nu } \\ 
B_{u\mu } & D_{\mu \nu }^{-1}
\end{array}
\right) \left( 
\begin{array}{l}
\Phi _{2} \\ 
A_{\nu }
\end{array}
\right) -\overline{c}D_{ug}^{-1}c,
\end{equation}
where the inverse propagators $D_{H}^{-1}$ and $D_{\mu \nu }^{-1}$ are given
in Eqs.(8) and (10) and $D_{uG}^{-1},$ $B_{u\mu }$ and $D_{ug}^{-1}$ are
given by 
\begin{equation}
D_{uG}^{-1}=p^{2}+m_{G}^{2}+\xi \text{ }u^{2}m_{A}^{2},
\end{equation}
\begin{equation}
B_{u\mu }=(1-u)m_{A}p_{\mu },
\end{equation}
and 
\begin{equation}
D_{ug}^{-1}=p^{2}+\xi u\text{ }m_{A}^{2}.
\end{equation}
As in case of the $R_{\xi }$ gauge, let us define $X_{u\mu \nu }^{-1}$ as 
\begin{equation}
X_{u\mu \nu }^{-1}\equiv D_{\mu \nu }^{-1}+B_{u\mu }D_{G}^{u}B_{u\nu
}=(p^{2}+m_{A}^{2})(\delta _{\mu \nu }-\frac{p_{\mu }p_{\nu }}{p^{2}})+\frac{%
D_{u}(p^{2})}{m_{G}^{2}+\xi \text{ }u^{2}m_{A}^{2}}\frac{p_{\mu }p_{\nu }}{%
\xi p^{2}},
\end{equation}
where 
\begin{equation}
D_{u}(p^{2})=p^{4}+(m_{G}^{2}\text{ }+2\text{ }u\xi m_{A}^{2})p^{2}+\xi
m_{A}^{2}(m_{G}^{2}+\xi u^{2}m_{A}^{2})\equiv
(p^{2}+m_{u+}^{2})(p^{2}+m_{u-}^{2}),
\end{equation}
to obtain the one-loop effective potential $V_{1}$as 
\begin{equation}
V_{1}=-\frac{\hbar }{2}Tr\ln D_{H}^{-1}-\frac{\hbar }{2}Tr\ln D_{uG}^{-1}
-\frac{\hbar }{2}Tr\ln
X_{u}^{-1}+\hbar \text{ }Tr\ln D_{ug}^{-1}.
\end{equation}
Again, by performing the one-loop momentum integral in $D\equiv
4-2\varepsilon $ dimension, we can obtain the renormalized one-loop
effective action in the MS scheme as 
\begin{eqnarray}
V &=&V_{0}+V_{1}=\frac{1}{2}m^{2}(\phi +v)^{2}+\frac{1}{24}\lambda (\phi
+v)^{4}+\frac{\hbar }{16\pi ^{2}}\{\frac{1}{4}m_{H}^{4}(\overline{\ln }\text{
}m_{H}^{2}-\frac{3}{2})+\frac{1}{4}m_{u+}^{4}(\overline{\ln }\text{ }%
m_{u+}^{2}-\frac{3}{2})+\frac{1}{4}m_{u-}^{4}(\overline{\ln }\text{ }%
m_{u-}^{2}-\frac{3}{2})  \nonumber \\
&&+\frac{3}{4}m_{A}^{2}(\overline{\ln }m_{A}^{2}-\frac{5}{6})-\frac{1}{2}{%
\xi }^{2}u^{2}m_{A}^{4}(\overline{\ln }\text{ }{\xi }um_{A}^{2}-\frac{3}{2}%
)\}.
\end{eqnarray}
The renormalization of the wave function for the scalar field can be done by
extracting the $\frac{1}{\epsilon }$ pole term from the Feynman diagram 
\begin{picture}(60,30) 
 \put(16,0){\line(1,0){8}} \put(32,0){\circle{16}} 
\put(5,-2) {$\Phi _{1}$}  \put(50,-2) {$\Phi _{1}$} 
\put(30,10) {$\Phi _{2}$} \put(40,0){\line(1,0){8}} 
\put(30,-15) {A} \end{picture}  by using the Feynman rules for the vertex $%
\Phi _{1}\Phi _{2}A_{\mu }$ given as 
\begin{equation}
\begin{picture}(80,20) \put(15,5){\vector(1,0){20}}
\put(35,5){\vector(2,1){20}} \put(35,5){\line(2,-1){20}} \put(0,2) {$\Phi
_{1}$} \put(60,15) {$\Phi _{2}$} \put(15,8) {$k_{1}$} \put(42,16) {$k_{2}$}
\put(60,-10) {$A_\mu$} \end{picture} =(1+u)g\text{ }k_{1\mu }+(1-u)g\text{ }%
k_{2\mu },
\end{equation}
and the propagators $D_{\mu \nu }$ and $D_{uG}$ given in Eqs.(10) and (33)
for the Goldstone field $\Phi _{2}$ and the photon field $A_{\mu }$. As a
result, we obtain 
\begin{equation}
\Phi _{1B}=\{1+\frac{\hbar }{2\epsilon (4\pi )^{2}}(3-\xi +2u\text{ }\xi
)g^{2}+\cdot \cdot \cdot \}\Phi _{1},,
\end{equation}
and obtain the one-loop gamma function as 
\begin{equation}
\gamma _{u\phi }=\frac{\hbar }{(4\pi )^{2}}(3-\xi +2u\text{ }\xi
)g^{2}+\cdot \cdot \cdot ,
\end{equation}
which agrees with that of the Fermi gauge when $u=0$ \cite{RG-1} and that of
the $\overline{R_{\xi }}$ gauge\cite{Kas} when $u=1$. Then, one can check
that the RG equation given in Eq.(19) is satisfied with $\beta _{\lambda }$
and $\beta _{m^{2}}$ given in Eq.(20) and with $\gamma _{v}=\gamma _{u\phi }$%
. As in case of $R_{\xi }$ gauge, we can obtain $v_{0}=\sqrt{-\frac{6m^{2}}{%
\lambda }}$ and 
\begin{eqnarray}
v_{1} &=&\frac{1}{32\pi ^{2}m^{2}}[\text{ }\frac{1}{2}m_{H}^{2}\frac{%
\partial m_{H}^{2}}{\partial \phi }(\overline{\ln }\text{ (}m_{H}^{2})-1)+%
\frac{1}{2}m_{u+}^{2}\frac{\partial m_{u+}^{2}}{\partial \phi }(\overline{%
\ln }\text{ (}m_{u+}^{2})-1)+\frac{1}{2}m_{u-}^{2}\frac{\partial m_{u-}^{2}}{%
\partial \phi }(\overline{\ln }\text{ (}m_{u-}^{2})-1)  \nonumber \\
&&+\frac{3}{2}m_{A}^{2}\frac{\partial m_{A}^{2}}{\partial \phi }(\overline{%
\ln }m_{A}^{2}-\frac{1}{3})-{\xi }^{2}u^{2}m_{A}^{2}\frac{\partial m_{A}^{2}%
}{\partial \phi }(\overline{\ln }\text{ }{\xi }um_{A}^{2}-1)\text{ ]}_{\phi
=0,\text{ }v=v_{0}}.
\end{eqnarray}
From Eq.(37), by writing $m_{u\pm }^{2}=p_{u}\pm \sqrt{q_{u}}$ we have 
\begin{equation}
q_{u}=m_{G}^{4}+4\xi (u-1)m_{A}^{2}m_{G}^{2}.
\end{equation}
Then one can see that $[\sqrt{q_{u}}$]$_{\phi =0,\text{ }v=v_{0}}=0$ and $[%
\frac{\partial q_{u}}{\partial \phi }]_{\phi =0,\text{ }v=v_{0}}\neq 0$ when 
$u\neq 1$ or $\xi \neq 0$ and as a result, $[\frac{\partial m_{u\pm }^{2}}{\partial
\phi }]_{\phi =0,\text{ }v=v_{0}}$ diverges. Although $[\frac{\partial m_{u\pm }^{2}}
{\partial \phi }]_{\phi =0,\text{ }v=v_{0}}$
 diverges, since $[m_{u+}^{2}]_{\phi =0,\text{ }%
v=v_{0}}=[m_{u-}^{2}]_{\phi =0,\text{ }v=v_{0}}=u\xi g^{2}v_{0}^{2},$ these
divergence contained in $\frac{1}{2}m_{u+}^{2}\frac{\partial m_{u+}^{2}}{%
\partial \phi }(\overline{\ln }$ ($m_{u+}^{2})-1)+\frac{1}{2}m_{u-}^{2}\frac{%
\partial m_{u-}^{2}}{\partial \phi }(\overline{\ln }$ ($m_{u-}^{2})-1)$
terms of Eq.(43) cancels out and the one-loop VEV $v_{1}$ converges for all
values of $u$. In case of $\overline{R_{\xi }}$ gauge we obtain
\begin{equation}
v_{1}=-\frac{v_{0}}{32\pi ^{2}}\{\lambda (\overline{\ln }(-2m^{2})-1)+\xi
g^{2}(\overline{\ln }(-\frac{6\xi g^{2}m^{2}}{\lambda })-1)+\frac{18g^{4}}{%
\lambda }(\overline{\ln }(-\frac{6g^{2}m^{2}}{\lambda })-\frac{1}{3})\},
\end{equation}
so that 
\begin{equation}
\mu \frac{\partial v_{1}}{\partial \mu }=\frac{1}{16\pi ^{2}}(\lambda +\xi
g^{2}+\frac{18g^{4}}{\lambda })v_{0}.
\end{equation}
By using Eq.(20) we have 
\begin{equation}
\mu \frac{dv_{0}}{d\mu }=\frac{\hbar }{16\pi ^{2}}(-\lambda +3g^{2}-\frac{%
18g^{4}}{\lambda })v_{0},
\end{equation}
and hence up to $O(\hbar )$, we obtain 
\begin{equation}
\mu \frac{dv}{d\mu }=\frac{\hbar }{16\pi ^{2}}(3+\xi )g^{2}v.
\end{equation}
which is consistent with Eq.(42) when $u=1.$ Finally, consider the IR
divergence of the VEV in case of the two-loop effective potential. Since the
mass of the Goldstone boson given by $m_{G}^{2}+{\xi }m_{A}^{2}$ in $%
\overline{R_{\xi }}$ gauge does not vanish in the limit $v\rightarrow v_{0}$
and $\phi \rightarrow 0$ when $\xi \neq 0,$ the term 
\begin{equation}
m_{X}^{2}\text{ }(m_{G}^{2}+{\xi }m_{A}^{2})\text{ }\overline{\ln }m_{X}%
\text{ }\overline{\ln }(m_{G}^{2}+{\xi }m_{A}^{2})
\end{equation}
coming from the Feynman diagram given in Eq.(27) in two-loop effective
potential and the resulting VEV coming from $[$ $\frac{\partial V_{2}}{%
\partial \phi }$]$_{\phi =0,\text{ }v=v_{0}}$ which can give the IR
divergence in case of the Landau gauge ($\xi =0)$ where Goldstone boson is
massless\cite{IR-1}\cite{IR-2} do not have an IR divergence as long as $\xi
\neq 0.$

In this paper, we have investigated the RG function and the perturbative
expansion of the VEV of the Abelian Higgs model th in $R_{\xi }$ and the
general gauge which extrapolate between the Fermi and the $\overline{R_{\xi }%
\text{ }}$ gauge by requiring the RG invariance of one-loop effective
potential. In case of the $R_{\xi }$ gauge, the gamma function of the VEV
that satisfy the RG equation for the effective potential was different from
the gamma function of the scalar field. In case of the general gauge, the
gamma function of the VEV obtained from the RG invariance of the effective
potential was same as that of the scalar field. When $u=1$ which corresponds
to the $\overline{R_{\xi }}$ gauge and $\xi =0$ which corresponds to Landau
gauge in general gauge, the one-loop VEV obtained from the
no-tadpole condition do not have the IR divergence and give correct RG
behavior. Both in $R_{\xi }\text{ }$ and $\overline{R_{\xi }\text{ }}$
gauge, the two-loop VEV have an IR divergence in Landau gauge($\xi =0$) and
these IR divergence do not occur when $\xi \neq 0$.

\end{document}